# Sustainability criterion implied externality pricing for resource extraction

Daniel Grainger

April 12, 2023

**Abstract**

A dynamic model is constructed that generalises the Hartwick and Van Long (2020) endogenous discounting setup by introducing externalities and asks what implications this has for optimal natural resource extraction with constant consumption. It is shown that a modified form of the Hotelling and Hartwick rule holds in which the externality component of price is a specific function of the instantaneous user costs and cross price elasticities. It is demonstrated that the externality adjusted marginal user cost of remaining natural reserves is equal to the marginal user cost of extracted resources invested in human-made reproducible capital. This lends itself to a discrete form with a readily intuitive economic interpretation that illuminates the stepwise impact of externality pricing on optimal extraction schedules.

**Keywords** externalities, sustainability, natural resources, optimal extraction, Hotelling's rule, Hartwick's rule

**JEL classification** D62, O44, Q01, Q20, Q32

## 1 Introduction

Hartwick and Van Long (2020) provide an elegant formulation of constant consumption sustainability with endogenous discounting. This is inspired by literature on endogenous discounting (see Le Kama and Schubert (2007), Obstfeld (1990) and Epstein (1987)) and literature pertaining to sustainability definitions that align with the Rawlsian max-min criterion (see Solow (1974), Asheim (2010), Cairns and Van Long (2006) and Mitra et al. (2013)). Hartwick and Van Long (2020) abstract away from externalities but note the importance of this consideration when referring to d'Autume et al. (2010) which has in turn built upon externality considerations of Stollery (1998). Whilst these contributions to the literature on externalities consider the implications of externalities impacting individual utility, they do not go so far as to assess externalities' effect on optimal extraction under constant consumption.

This paper starts out staying deliberately close to Hartwick and Van Long (2020) paying due regard to this, the closest and arguably only piece of literature on the specific question of how a constant consumption sustainability criterion relates to externalities, and then diverges into exploring externalities and their implications on Hotelling, Hartwick and optimal extraction rules generally. The elasticity concept is identified as central to modelling externalities in a setting of numerous natural resources and employed in a similar manner to the approach of Hartwick (1978). It is shown that an externalities' Hartwick and Hotelling rule (Hartwick, 1977; Hotelling, 1931) is applicable in a constant consumption world with endogenous discounting. This then implies that the externality adjusted price of remaining natural reserves is equal to the return on extracted resources invested in human-made reproducible capital and hence how externality pricing affects the evolution of optimal extraction schedules.

## 2 The basic model

The Hartwick and Van Long (2020) model is now generalised to allow for externalities. Infinitely lived individuals are indexed by $\theta \in [0,1]$ and we assume zero population growth. Strictly increasing and concave heterogeneous individual utility functions exist; where at time $t$ with consumption $c(t,\theta)$ the utility of the individual $\theta$ is denoted $u_\theta\big(c(t,\theta)\big)$. Capital stock and natural resource stocks are owned by an individual and are respectively denoted $k(t,\theta)$ and $x_j(t,\theta)$ for various resource stocks indexed by $j \in \mathbb{N}$. Resource flows due to an individual extracting from stock $x_j(t,\theta)$ is denoted $q_j(t,\theta)$ with the dynamics of this extraction described by $\dot{x}_j(t,\theta) = G_j\big(x_j(t,\theta)\big) - q_j(t,\theta)$ where $G_j\big(x_j(t,\theta)\big)$ is the natural resource growth and $\dot{x}_j(t,\theta)$ denotes the first derivative with respect to time of $x_j(t,\theta)$. Note that if $G_j\big(x_j(t,\theta)\big) \not> 0$ the resource is not renewable. The price of resource $j$ at time $t$ is denoted $p_j(t)$ and the rate of return on the capital stock is denoted $r(t)$. The individual's income and capital accumulation are respectively $y(t,\theta) = r(t)k(t,\theta) + \sum_j p_j(t)q_j(t,\theta)$ and $\dot{k}(t,\theta) = y(t,\theta) - c(t,\theta)$.

At the aggregate level the economy's capital stock, resource stock $j$, resource flows, consumption and income are respectively $K(t) = \int_0^1 k(t,\theta)d\theta$, $X_j(t) = \int_0^1 x_j(t,\theta)d\theta$, $Q_j(t) = \int_0^1 q_j(t,\theta)d\theta$, $C(t) = \int_0^1 c(t,\theta)d\theta$ and $Y(t) = \int_0^1 y(t,\theta)d\theta$. The inverse demand function of a natural resource is assumed to be strictly monotonic with respect to extraction of resources with $\frac{\partial p_j(t)}{\partial Q_k(t)}$ well behaved for all $k$.

### 2.1 Individual optimisation with externalities

Each individual maximizes the present value of utility flows by choosing paths for their consumption $c(t,\theta)$ and extraction for each resource $q_j(t,\theta)$ i.e., $max \int_0^\infty u_\theta\big(c(t,\theta)\big)\beta(t)d\theta$ where $\beta(t)$ is the endogenous discount factor of Hartwick and Van Long (2020).

For notational simplicity the dependence on $t$ and $\theta$ will be suppressed in the following.

Thus, the maximisation becomes $max \int_0^\infty u\beta \, d\theta$ which is subject to constraints

$$\dot{k} = rk + \sum_j p_j q_j - c, \forall j: \dot{x}_j = G_j(x_j) - q_j, c \geq 0, q_j \geq 0, \lim_{t \to \infty} x_j \geq 0, \lim_{t \to \infty} k \geq 0$$

The associated Hamiltonian with state variables $x_j, k$ and costate variables $\pi, \psi_j$ is $H = u\beta + \pi\big(rk + \sum_j p_j q_j - c\big) + \sum_j \psi_j\big(G_j(x_j) - q_j\big)$ for which the interior solutions are of interest and imply optimality conditions

$$\forall j: \pi\left(p_j\left(1 + \sum_k \frac{1}{\varepsilon_{jk}}\frac{p_k q_k}{p_j q_j}\right)\right) - \psi_j = 0 \text{ where } \varepsilon_{jk} = \frac{\partial q_j}{\partial p_k}\frac{p_k}{q_j} \text{ is the cross-price elasticity}$$

$$u'\beta - \pi = 0$$

$$\dot{\pi} = -\frac{\partial H}{\partial k} = -\pi r$$

$$\forall j: \dot{\psi}_j = -\frac{\partial H}{\partial x_j} = -\psi_j G'_j$$

$$\lim_{t \to \infty} \pi k = 0 \text{ and } \forall j: \lim_{t \to \infty} \psi_j x_j = 0$$

The first two equations of the above optimality conditions imply the following:

$ln(u') + ln(\beta) = ln(\pi)$ and $\forall j: ln(\pi) + ln(p_j) + ln\left(1 + \sum_k \frac{1}{\varepsilon_{jk}} \frac{p_k q_k}{p_j q_j}\right) = ln(\psi_j)$ which when differentiated with respect to time becomes $\frac{u''\dot{c}}{u'} + \frac{\dot{\beta}}{\beta} = \frac{\dot{\pi}}{\pi}$ and $\forall j: \frac{\dot{\pi}}{\pi} + \frac{\dot{p}_J}{p_j} + \frac{\overline{\left(1+\sum_k \frac{1}{\varepsilon_{jk}} \frac{p_k q_k}{p_j q_j}\right)}}{\left(1+\sum_k \frac{1}{\varepsilon_{jk}} \frac{p_k q_k}{p_j q_j}\right)} = \frac{\dot{\psi}_J}{\psi_j}$

Applying the last two equations of the above optimality conditions and making the constant consumption assumption implies that $\frac{\dot{\beta}}{\beta} = -r$ and $\forall j: -r + \frac{\dot{p}_J}{p_j} + \frac{\overline{\left(1+\sum_k \frac{1}{\varepsilon_{jk}} \frac{p_k q_k}{p_j q_j}\right)}}{\left(1+\sum_k \frac{1}{\varepsilon_{jk}} \frac{p_k q_k}{p_j q_j}\right)} = -G'_j$

This may be rearranged into the externality modified Hotelling rule $\forall j: \frac{\overline{p_J\left(1+\sum_k \frac{1}{\varepsilon_{jk}} \frac{p_k q_k}{p_j q_j}\right)}}{p_j\left(1+\sum_k \frac{1}{\varepsilon_{jk}} \frac{p_k q_k}{p_j q_j}\right)} = r - G'_j$ the interpretation of which is the rate of increase in externality adjusted resource price is equal to the interest rate net of the growth rate of that natural resource. Notice that the additional externality price is represented here by $\left(\sum_k \frac{p_k q_k}{\varepsilon_{jk} p_j q_j}\right) p_j$ where the bracketed term may be interpreted as externalities' price margin which can push the price away from the perfectly competitive equilibrium price $p_j$ thereby inducing a welfare cost. This margin is the summation of relative instantaneous user costs accounting for cross price elasticity effects where the percentage change in the price of $p_k$ causes a percentage change in quantity $q_j$. Notice that for perfect elasticities, in the limit, this bracketed term becomes zero and the modified Hotelling rule becomes the case derived in Hartwick and Van Long (2020). Further restriction of no growth reduces the modified Hotelling rule to the original Hotelling (1931) rule. Given $\varepsilon_{jk} > 0$ indicates a substitution effect and $\varepsilon_{jk} < 0$ indicates a complement effect, a net substitution effect reflected in the externalities' price margin is associated with a resource price incorporating externalities such that $p_j \left(1 + \sum_k \frac{1}{\varepsilon_{jk}} \frac{p_k q_k}{p_j q_j}\right) > p_j$; which agrees with the intuition that an externality price in excess of the perfect competition price reflects foregone substitution opportunities with an efficient market for externalities forming the associated price $\left(\sum_k \frac{p_k q_k}{\varepsilon_{jk} p_j q_j}\right) p_j$.

**Proposition 1:** *a modified form of the Hotelling rule holds in which the externality component of price is a specific function of the instantaneous user costs and cross price elasticities*

## 2.2 The equilibrium with externalities

A benevolent social planner is assumed to maximise constant aggregate consumption $\bar{C}$ i.e., max $\bar{C}$ subject to constraints $\dot{K} = Y - \bar{C}$, $\forall j: \dot{X}_J = G_j(X_j) - Q_j$, $Q_j \geq 0$, $\lim_{t \to \infty} X_j \geq 0$ and $\lim_{t \to \infty} K \geq 0$.

The associated Hamiltonian is therefore $H = \pi^S(Y - \bar{C}) + \sum_j \psi^S{}_j \left(G_j(X_j) - Q_j\right) = 0$. Given aggregate income net of consumption is invested, the Hamiltonian may be rearranged such that investment $I = Y - \bar{C} = \sum_j \frac{\psi^S{}_j}{\pi^S}\left(Q_j - G_j(X_j)\right)$.

Notice that $\forall j: \pi\left(p_j\left(1 + \sum_k \frac{1}{\varepsilon_{jk}} \frac{p_k q_k}{p_j q_j}\right)\right) - \psi_j = 0$ established earlier may be rewritten as $\forall j: \pi\left(p_j q_j + \sum_k \frac{p_k q_k}{\varepsilon_{jk}}\right) - \psi_j q_j = 0$ and upon integrating across individuals $\forall j: \int_0^1 \pi\left(p_j q_j + \sum_k \frac{p_k q_k}{\varepsilon_{jk}}\right) d\theta = \int_0^1 \psi_j q_j d\theta$ we arrive at $\pi\left(p_j Q_j + \sum_k \frac{p_k Q_k}{\varepsilon_{jk}}\right) - \psi_j Q_j = 0$. Given the costate variables $\psi^S{}_j$ and $\pi^S$ are the shadow prices for the constraints identical in value to $\psi_j$ and $\pi$ then $\frac{\psi^S{}_j}{\pi^S} = \frac{\psi_j}{\pi} = p_j\left(1 + \sum_k \frac{1}{\varepsilon_{jk}} \frac{p_k Q_k}{p_j Q_j}\right)$ and therefore $I = \sum_j \left[p_j\left(1 + \sum_k \frac{1}{\varepsilon_{jk}} \frac{p_k Q_k}{p_j Q_j}\right)\left(Q_j - G_j(X_j)\right)\right]$; which is the modified Hartwick rule accounting for externalities at the industry level for resource $j$. Notice that for perfect elasticities $\sum_k \frac{1}{\varepsilon_{jk}} \frac{p_k Q_k}{p_j Q_j}$ becomes zero and the modified Hartwick rule becomes the case derived in Hartwick and Van Long (2020). Further restriction of no natural resource growth reduces the modified Hartwick rule to the original Hartwick (1977) rule.

**Proposition 2:** *a modified form of the Hartwick rule holds in which the externality component of price is a specific function of the instantaneous user costs and cross price elasticities*

### 3 Resource optimal extraction rule for constant consumption with externalities

The previously established $\frac{\psi_j}{\pi} = p_j\left(1 + \sum_k \frac{1}{\varepsilon_{jk}} \frac{p_k q_k}{p_j q_j}\right)$ motivates an exploration into the relationship between shadow prices of the natural resource reserves and the related instantaneous user costs facing the individual as follows. Simplifying the expression to $\psi_j = \pi \frac{\partial(\sum_k p_k q_k)}{\partial q_j}$ and differentiating with respect to time arrives at $\frac{d}{dt}(\psi_j) = \pi \frac{d}{dt}\left(\frac{\partial(\sum_k p_k q_k)}{\partial q_j}\right) + \frac{d}{dt}(\pi) \frac{\partial(\sum_k p_k q_k)}{\partial q_j}$. Because of the Hamiltonian specification and since $\dot{x}_j = G_j - q_j$, then the Euler LaGrange relationship $\frac{d}{dt}\left(\frac{\partial(\sum_k p_k q_k)}{\partial q_j}\right) = -\frac{\partial(\sum_k p_k q_k)}{\partial x_j}$ holds. By substitution $\frac{d}{dt}(\psi_j) = -\pi \frac{\partial(\sum_k p_k q_k)}{\partial x_j} + \frac{d}{dt}(\pi) \frac{\partial(\sum_k p_k q_k)}{\partial q_j}$ results. Substituting $\forall j: \dot{\psi}_j = -\psi_j G'_j$ results in $-\psi_j G'_j = -\pi \frac{\partial(\sum_k p_k q_k)}{\partial x_j} + \dot{\pi} \frac{\partial(\sum_k p_k q_k)}{\partial q_j}$ and further substitution of $\dot{\pi} = -\pi r$ results in $\psi_j G'_j = \pi \frac{\partial(\sum_k p_k q_k)}{\partial x_j} + \pi r \frac{\partial(\sum_k p_k q_k)}{\partial q_j}$. Combining this with $\psi_j = \pi \frac{\partial(\sum_k p_k q_k)}{\partial q_j}$ allows for the removal of costates $\pi$ and $\psi_j$ in the expression and $\frac{\partial(\sum_k p_k q_k)}{\partial q_j} G'_j = \frac{\partial(\sum_k p_k q_k)}{\partial x_j} + r \frac{\partial(\sum_k p_k q_k)}{\partial q_j}$ results. Finally, rearranging this arrives at $-\frac{\partial(\sum_k p_k q_k)}{\partial x_j} = (r - G'_j) \frac{\partial(\sum_k p_k q_k)}{\partial q_j}$ which states that the marginal value of remaining reserves $x_j$ equals the marginal value of extracted resources invested in reproducible capital at the financial rate of return net of the resource growth rate on the optimal extraction path. Notice that the left-hand side term of the equation is the derivative of the user costs with respect to the stock of natural resource reserves i.e., it is a measure of the opportunities forgone per unit of reserves inclusive of externalities called the marginal user cost of reserves. The right-hand side final term is similarly the derivative of the user costs with respect to the extracted quantity of natural resources i.e., it is a measure of the opportunities forgone per unit of extracted resources called the marginal user cost of extracted resources inclusive of externalities called the marginal user cost of extracted resources.

Now, rewrite as $-\frac{\partial(\sum_k p_k q_k)}{\partial x_j} = \left(1 + r - (1 + G'_j)\right)\frac{\partial(\sum_k p_k q_k)}{\partial q_j}$ and rearrange to obtain

$\frac{\left(-\frac{\partial(\sum_k p_k q_k)}{\partial x_j}\frac{1}{(1+G'_j)} + \frac{\partial(\sum_k p_k q_k)}{\partial q_j}\right)(1+G'_j)}{(1+r)} = \frac{\partial(\sum_k p_k q_k)}{\partial q_j}$ and notice that the terms in the first bracket of the numerator on the left hand side of the equation relate to the natural resource with growth reversed out of the first term which then allows natural resource growth to be capture in the second bracket of the numerator. The denominator of the left-hand side converts the numerator to a present value. The right-hand side of the equation is the present value of the extracted resource all of which may be interpreted in the following proposition which aligns with the textbook-type dynamic extraction marginal argument but rigorously established here with a precise form incorporating natural resources' growth and externalities implicit in prices.

**Proposition 3:** for constant consumption on the optimal extraction path with externalities, *the present value of the marginal user cost of natural resources available with possible natural resource growth for later extraction is equal to the present value of the marginal user cost of extracted resources invested in capital stock with possible human-made capital growth.*

A discretised version of the above provides insight into the stepwise evolution of the optimal path with constant consumption and prices reflecting the marginal user costs inclusive of externalities. To simplify into a discrete world let $P_t$ denote the price of a unit available for extraction at time $t$ and $P_{t+\Delta t}$ denote the price of a unit available for extraction at time $t + \Delta t$, then each unit at $t$ available for later extraction grows to $1 + G'_j$ with a value of $P_{t+\Delta t}(1 + G'_j)$ at $t + \Delta t$ and each unit extracted at $t$ valued at $P_t$ and invested in the capital stock grows to $P_t(1 + r)$ at $t + \Delta t$. By proposition 3, $\frac{P_{t+\Delta t}(1+G'_j)}{(1+r)} = P_t$. This has implications for user costs. Consider total reserves $X_t^{(T)}$ at $t$ for which natural resource extraction of $Q_t$ at $t$ has a user cost at $t + \Delta t$ of $P_t Q_t(1 + r)$ with the remaining resources $X_t^{(T)} - Q_t$ at $t$ growing to a user cost at at $t + \Delta t$ of $P_{t+\Delta t}(1 + G'_j)(X_t^{(T)} - Q_t)$. Notice that $Q_t$ in $P_t Q_t(1 + r)$ and $P_{t+\Delta t}(1 + G'_j)(X_t^{(T)} - Q_t)$ may be varied to a value at which $P_t Q_t(1 + r) = P_{t+\Delta t}(1 + G'_j)(X_t^{(T)} - Q_t)$, called the user cost rule, and may be considered to hold in the neighbourhood of some $Q_t$. Because the user cost rule with small perturbations in the neighbourhood of $Q_t$ is assumed to hold, an arbitrarily small additional $\Delta Q_t$ units of extracted resource at $t$ and an arbitrarily small additional $\Delta Q_t$ units of resources in reserve at $t$ implies that $P_t(Q_t + \Delta Q_t)(1 + r) = P_{t+\Delta t}(1 + G'_j)(Q_t^{(T)} - Q_t + \Delta Q_t)$. Combining this implication with the assumption that $P_t Q_t(1 + r) = P_{t+\Delta t}(1 + G'_j)(Q_t^{(T)} - Q_t)$ holds and subtracting one from the other results in the proposition 3 implied optimality path condition $\frac{P_{t+\Delta t}(1+G'_j)}{(1+r)} = P_t$. That is, the user cost rule ensures $Q_t$ is on the optimal path.

**Proposition 4:** in a discretised setting for constant consumption and prices inclusive of externalities on the optimal extraction path, it is a sufficient condition that the user cost rule holds i.e., "*the present value of the user cost of natural resources available with possible natural resource growth for later extraction is equal to the present value of the user cost of extracted resources invested in capital stock with possible human-made capital growth*".

## 4 Concluding remarks

It has been shown in this paper that generalising Hartwick and Van Long (2020) to allow for externalities leads to externality modified versions of Hotelling (1931) and Hartwick (1977) rules that depend on instantaneous user costs and cross price elasticities. Furthermore, on a constant consumption optimal path, the externality adjusted marginal user cost of remaining natural reserves is equal to the marginal user cost of extracted resources invested in human-made reproducible capital. A sufficient condition for this is to ensure that, at any instant, the present value of the user cost of natural resources available with possible natural resource growth for later extraction is equal to the present value of the user cost of extracted resources invested in capital stock with possible human-made capital growth with marginal user costs incorporating the externalities' price. The approach taken here offers a different perspective of externalities than that of d'Autume et al. (2010) and notably not requiring of Cobb Douglas production and utility assumptions. Specifically, externalities in the present paper are not restricted to enter directly via the individual utility function and instead are seen through their effect i.e., deviations from perfectly competitive resource prices. The modelling here abstracts away from monopolistic behaviour of which a comprehensive treatment is obviously provided by Hotelling (1931) albeit without explicit regard to externalities. Preferring simplicity in the derivation here and as was done in Hartwick (1977) a zero population growth is assumed in the modelling; something which may be relaxed in future work possibly drawing inspiration from Asheim et al. (2021) and Asheim et al. (2023).